\documentclass[twocolumn]{aastex62}
\begin{document}
\title{H$^-$ Opacity and Water Dissociation in the Dayside Atmosphere of the Very Hot Gas Giant WASP-18~\MakeLowercase{b}}

\author{Jacob Arcangeli}
\affiliation{Anton Pannekoek Institute for Astronomy, University of Amsterdam, Science Park 904, 1098 XH Amsterdam, The Netherlands }, 
\author{Jean-Michel D\'{e}sert} 
\affiliation{Anton Pannekoek Institute for Astronomy, University of Amsterdam, Science Park 904, 1098 XH Amsterdam, The Netherlands }, 
\author{Michael R. Line} 
\affiliation{School of Earth \& Space Exploration, Arizona State University, Tempe AZ 85287, USA }, 
\author{Jacob L. Bean} 
\affiliation{Department of Astronomy \& Astrophysics, University of Chicago, 5640 S. Ellis Avenue, Chicago, IL 60637, USA}, 
\author{Vivien Parmentier} 
\affiliation{Aix Marseille Univ, CNRS, LAM, Laboratoire d'Astrophysique de Marseille, Marseille, France },
\author{Kevin B. Stevenson} 
\affiliation{Space Telescope Science Institute, 3700 San Martin Drive, Baltimore, MD 21218, USA },
\author{Laura Kreidberg} 
\affiliation{Harvard-Smithsonian Center for Astrophysics, Cambridge, MA 02138, USA},
\affiliation{Harvard Society of Fellows, 78 Mt. Auburn St., Cambridge MA 02138},
\author{Jonathan J. Fortney} 
\affiliation{Department of Astronomy and Astrophysics, University of California, Santa Cruz, CA 95064},  
\author{Megan Mansfield}
\affiliation{Department of Geophysical Sciences, University of Chicago, 5734 S. Ellis Avenue, Chicago, IL 60637, USA},
\author{Adam P. Showman} 
\affiliation{Department of Planetary Sciences and Lunar and Planetary Laboratory, University of Arizona, Tucson, Arizona 85721, USA}

\begin{abstract}
We present one of the most precise emission spectra of an exoplanet observed so far. We combine five secondary eclipses of the hot Jupiter WASP-18~b (T$_{day}\sim2900$ K) that we secured between 1.1 and 1.7 $\mu$m with the WFC3 instrument aboard the Hubble Space Telescope. Our extracted spectrum (S/N=50, R$\sim$40) does not exhibit clearly identifiable molecular features but is poorly matched by a blackbody spectrum. We complement this data with previously published Spitzer/IRAC observations of this target and interpret the combined spectrum by computing a grid of self-consistent, 1D forward models, varying the composition and energy budget. At these high temperatures, we find there are important contributions to the overall opacity from H$^-$ ions, as well as the removal of major molecules by thermal dissociation (including water), and thermal ionization of metals.
These effects were omitted in previous spectral retrievals for very hot gas giants, and we argue that they must be included to properly interpret the spectra of these objects.
We infer a new metallicity and C/O ratio for WASP-18~b, and find them well constrained to be solar ([M/H]$ = -0.01\pm 0.35$, C/O $<0.85$ at $3\sigma$ confidence level), unlike previous work but in line with expectations for giant planets. The best fitting self-consistent temperature-pressure profiles are inverted, resulting in an emission feature at 4.5 $\mu$m seen in the Spitzer photometry. These results further strengthen the evidence that the family of very hot gas giant exoplanets commonly exhibit thermal inversions.
\end{abstract}


\section{Introduction}

Hot Jupiters are the easiest exoplanets to study because they are hot enough to have most or all of their atmospheric constituents in gas phase.
From the growing number of known exoplanets, the population of very-hot hot Jupiters has emerged \citep{Sudarsky2000}. This subset spans a range of dayside temperatures from 2500-4600~K, with the hottest being as hot as the photosphere of a K-dwarf star (KELT-9b; \citealt{Gaudi2017}).
These extreme planets are currently being discovered by ground-based surveys that focus on bright stars. 
Several important questions have emerged from the study of these highly irradiated planets, including the influence of stellar irradiation on their inflated radii and mass loss rate, their atmospheric composition, and the frequency and origin of stratospheric thermal inversions.

\cite{Hubeny2003} first proposed the possibility of a bifurcation in the thermal structure of giant exoplanet atmospheres. Strong irradiation combined with efficient optical absorbers in the atmosphere (such as TiO and VO in gas-phase) could cause an inversion layer in the temperature-pressure profile \citep{Fortney2006,Fortney2008,Burrows2008,Parmentier2015}.

Recent observations of some of these extreme hot giants have revealed temperature inversions in their atmospheres (WASP-33~b: \citealt{Haynes2015}, WASP-121~b: \citealt{Evans2017}, WASP-18~b: \citealt{Sheppard2017}). Nevertheless, for all these studies, the retrieved metallicities and abundances are much higher than expected for a solar composition (e.g. VO 1000x solar for WASP-121~b, metallicity $\sim$300x solar with a C/O$\sim$1 for WASP-18~b). This is surprising for such massive gas giants, as their expected formation channels imply that their metallicities should be close to their host stars', as observed in their cooler counterparts (e.g., \citealt{Kreidberg2014b, Benneke2015, Line2016}).\par

In this paper, we argue that chemistry and opacity sources that are well known to operate at high temperatures from stellar astrophysics are key to the interpretation of very hot gas giant atmospheres. In particular, some of the primary sources of opacity on the daysides of these atmospheres will thermally dissociate at sufficiently low pressures and high temperatures. A second consideration is the effect of thermal ionization, whose electrons provide the seeds for bound-free and free-free interactions with atomic hydrogen that generate H$^-$ opacity (see Section \ref{sec:opacity}). While these effects are included in some models of very hot gas giants, in particular those that assume radiative-convective equilibrium, (e.g., \citealt{Barman2001, Burrows2008,Fortney2008}), their consequences for spectral retrieval have not yet been explored.

In this context, we present a new analysis and interpretation of observations obtained with the Hubble Space Telescope Wide Field Camera 3 (HST WFC3) and Spitzer Infrared Array Camera (IRAC) of the dayside emission spectrum of WASP-18~b.
WASP-18~b \citep{Hellier2009} is a 10 10M$_{J}$ planet that orbits a bright F6 host (Vmag=9.3) on a short period (0.94 days), and has an equilibrium temperature of 2700K. In Section~\ref{Sec:Data} we present the observations and data analysis. In Section~\ref{Sec:Results} we discuss the effect of thermal dissociation and H$^-$ opacities on the interpretation of this emission spectrum.

\section{Observations and Data Reduction}
\label{Sec:Data}
\subsection{Observations}
\par
Our team observed five secondary eclipses of WASP-18b with 24 orbits of the HST as part of a larger Treasury program (GO-13467), including a phase-curve presented in a separate paper (Arcangeli et al. in prep.). We concentrate here on the secondary eclipse observations. The data were obtained with HST/WFC3, with the G141 grism, covering 1.1 to 1.7$\mu$m, using the spatial scanning technique in both directions. Individual pixels in the spectrum reached a maximum flux level of 30,000 electrons, below 40\% of full-well saturation, where the pixel response is linear. \par
The first two eclipses were taken using the 256x256 pixel subarray (SPARS10, NSAMP=12, 74s exposures), however the spectrum was seen to leak outside of this subframe. Subsequent data used the 512x512 subarray (SPARS10, NSAMP=16, 112s exposures) with the same scan rate. We remove part of the second eclipse's final orbit, due to a loss of fine-guidance.

\subsection{Data reduction and analysis}
We developed a custom data reduction and analysis pipeline following the methods outlined in \cite{Kreidberg2014a}. We first form subexposures from each full exposure by subtracting consecutive non-destructive reads. 
We calibrate the wavelength solution using a direct image taken at the start of each visit. We apply a wavelength-dependent flat-field correction and remove cosmic rays using a local median filter. We calculate the average sky background by masking the spectra on each subexposure, iteratively clipping outlier pixels. We finally apply an optimal extraction algorithm \citep{Horne1986} to maximise the signal-to-noise from each subexposure.\par
The reduced light curves are dominated by time-dependent systematics characteristic of HST observations. We parametrise these using the {\fontfamily{cmr}\selectfont model-ramp} technique with a single exponential in time and a linear visit-long slope. We remove the first orbit of each visit from our analysis. The second orbit is parametrised with a separate ramp amplitude. We compare the {\fontfamily{cmr}\selectfont model-ramp} technique with a common-mode correction and find we reach the same precision in each of the light-curve fits. \par
We fit for the instrument systematics, the eclipse-depth, and eclipse time simultaneously for a total of 7 free parameters for each of the white light curves. We then bin the data into 14 wavelength channels and fit again in each channel whilst maintaining the ramp timescale and eclipse time fixed to the white-light curve values. The remaining system parameters are fixed to literature values from \cite{Southworth2009}. We combine the five extracted secondary eclipse spectra since we find that each are consistent within one sigma.
The residuals from the white light curve fits range from 1.05x to 2x the photon noise limit. The precision reached in the spectroscopic fits is less than 1.2x photon noise for each bin. \par
In order to estimate the errors on our fitted parameters and identify the degeneracies in the model we use a Markov chain Monte Carlo approach using the open-source EMCEE code \citep{emcee}.
We test convergence by employing the Gelmaan-Rubin diagnostic for each chain of 10,000 steps with 400 walkers. Our final precision on the spectroscopic eclipse depths is 20 ppm per wavelength bin in the combined spectrum, achieving a signal-to-noise ratio of 50 at a resolution of R$\sim40$, shown in Table \ref{tab:spec}. Our combined spectrum is consistent with \cite{Sheppard2017}.

\begin{figure*}
\begin{center}
\includegraphics[scale=0.35]{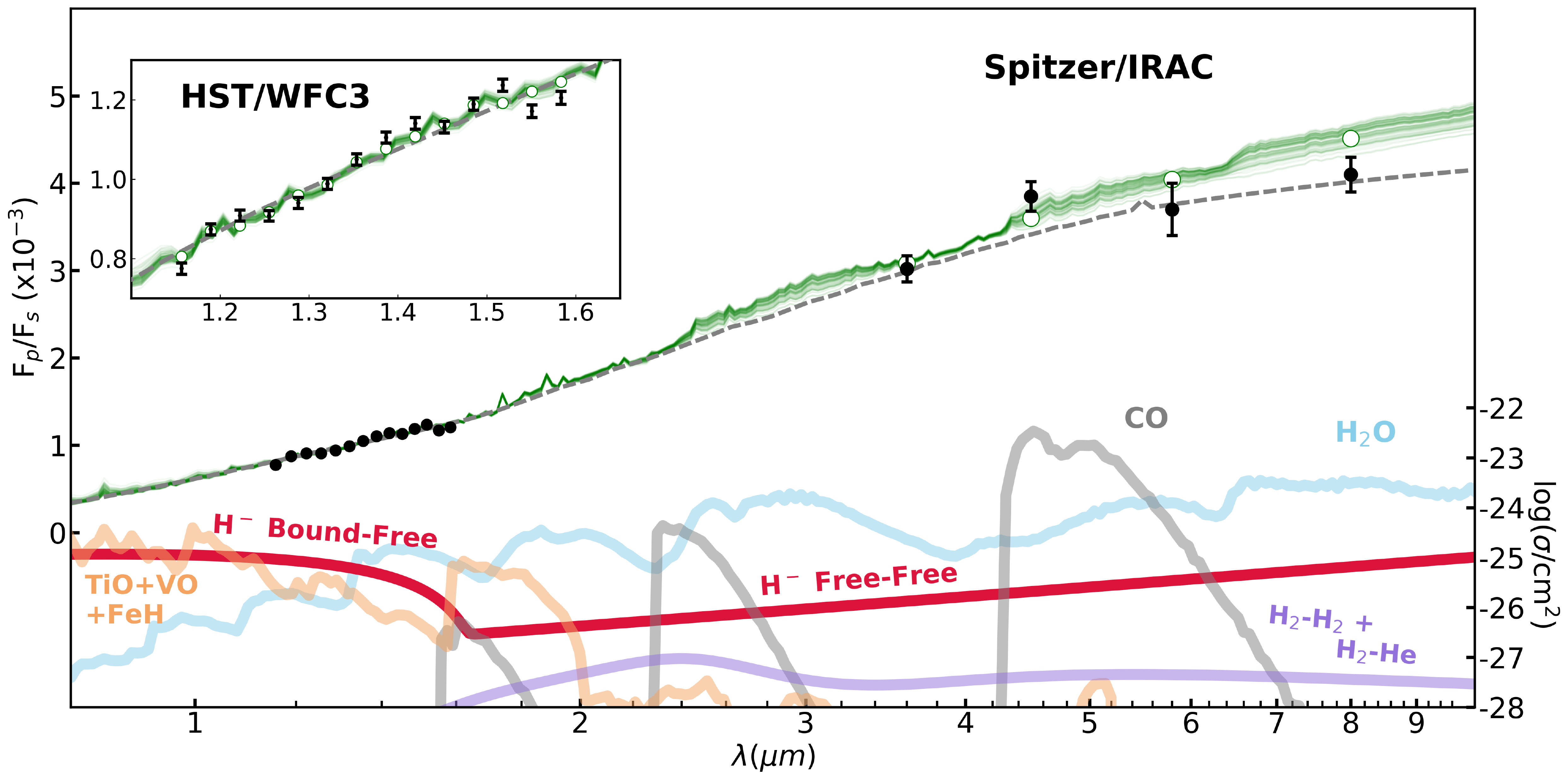}
\caption{Dayside eclipse spectrum (black points) from this work and previous Spitzer observations compared to best-fit model spectrum (white circles). The best-fit model has a reduced chi-squared of 2.0. In green are 100 samples from the posterior of the model spectrum derived from the grid retrieval, and in grey the best-fit blackbody spectrum to the WFC3 data of 2890$\pm$47~K. Dominant opacity cross sections weighted by their molecular abundances (log($\sigma$)) are shown for key molecules, taken at a pressure level of 0.33 bar (the peak of the WFC3 emission).}\label{fig:spec}
\end{center}
\end{figure*}

\begin{figure}
\begin{center}
\includegraphics[scale=0.35]{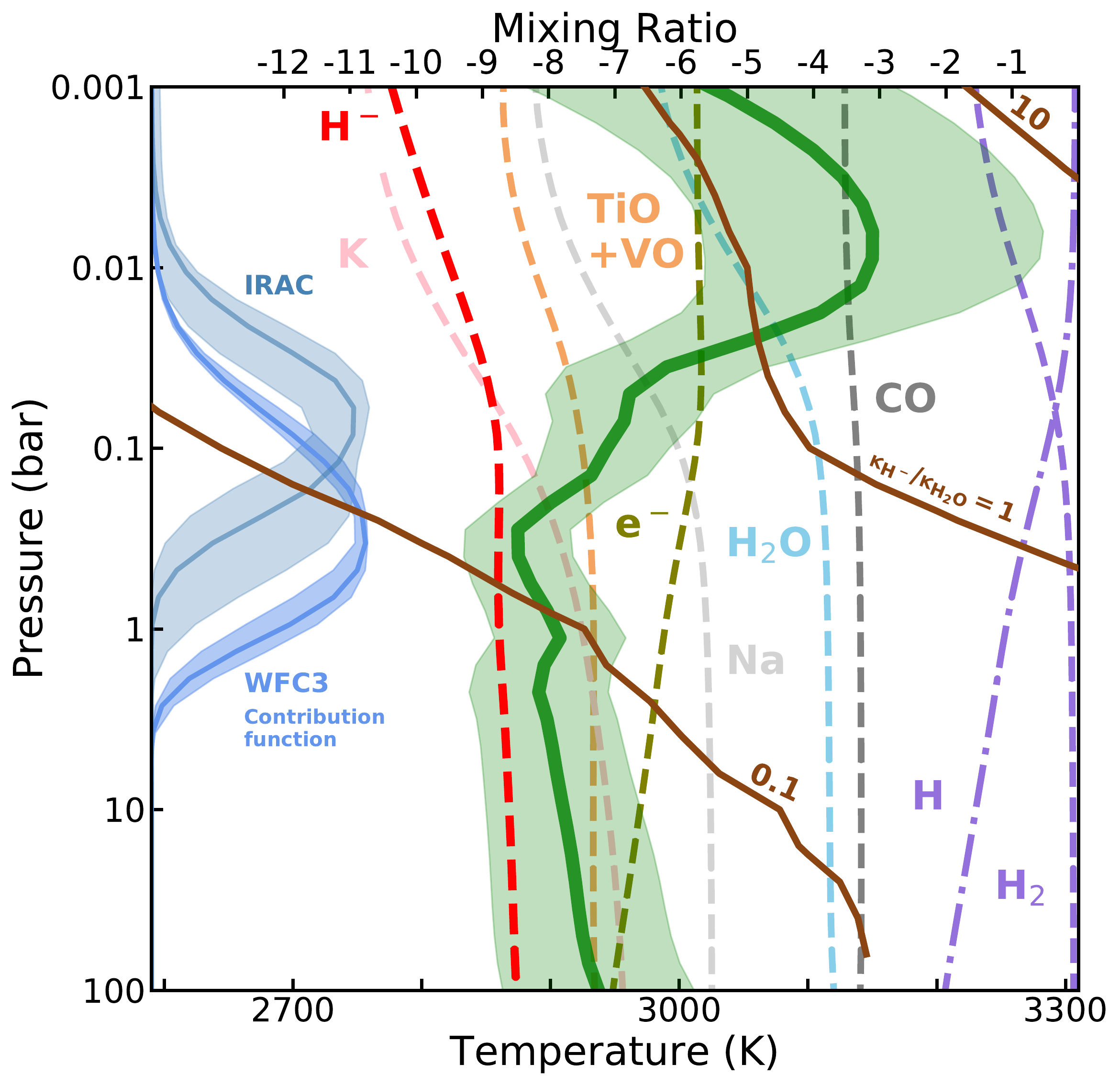}
\caption{Best fit T-P profile shown in green with 1$\sigma$ error contours. Flux contribution functions are shown on the left for HST/WFC3 and Spitzer/IRAC in blue with 1$\sigma$ regions. Dashed lines show the logarithm of the mixing ratios for key species at different pressures. Lines in brown denote contours of the ratio between the bound-free opacity of H$^-$ at 1.25$\mu$m and molecular gas opacity, mainly from H$_2$O and TiO. 
}\label{fig:spaghetti}
\end{center}
\end{figure}

\section{Results \& Discussion}
\label{Sec:Results}
\label{subsec:Discussion}
\par
The combined WFC3 emission spectrum (show in Figure \ref{fig:spec}) does not exhibit spectral features expected from molecules such as H$_2$O or TiO. We complement the WFC3 emission spectrum with four Spitzer/IRAC data points already published \citep{Nymeyer2011, Maxted2013}, after ensuring that the system parameters are consistent, and we present below several scenarios to explain this combined spectrum. \par

\subsection{Fitting a blackbody spectrum}
We first test whether the WFC3 emission spectrum is consistent with a simple blackbody spectrum, which would be caused by an isothermal atmosphere over the pressures probed. We find a best fit blackbody temperature of 2890$\pm 47$~K, using a PHOENIX stellar model of T=6400K, logg=4.5, and [M/H]=0.0 for the star. However, this is a relatively poor fit to the data, with a reduced $\chi ^2$ of 3.1.

The Spitzer/IRAC photometric points at 3.5, 5.8 and 8.0 $\mu$m lie on the blackbody spectrum extrapolated from our WFC3 data, but the planet's flux at 4.5 $\mu$m is larger by 2$\sigma$, suggesting the presence of emission features (see Figure \ref{fig:spec}). In this wavelength range the dominant opacity sources are CO and H$_2$O, and spectral features would appear in emission only if the temperature-pressure profile of the atmosphere were inverted, and not isothermal. However the lack of water spectral features at 1.4 ~$\mu$m could appear to be at odds with this conclusion. Previous studies have explained WASP-18b's spectrum with a high C/O ratio that can deplete the gas-phase water and remove its spectral features whilst allowing for a non-isothermal atmosphere \citep{Sheppard2017}. In the following section, we present an alternative explanation taking into account the key changes in opacity at these high temperatures, due to molecular dissociation, thermal ionization, and the presence of H$^-$ ions, while requiring nominal solar metallicity and C/O.

\par
\subsection{Opacity sources in very hot gas giant exoplanet atmospheres}
\label{sec:opacity}

The dominant opacity sources in the near infrared (NIR) for hot Jupiters are H$_2$O, CO, and metal hydrides and oxides. However, for the very-hot hot Jupiters (T$>2500$~K), a significant fraction of water also thermally dissociates at low pressures \citep{Parmentier2017}. In cool stellar photospheres with similar temperatures, large water absorption features can still be observed in their spectra as the increased pressure at the photosphere due to their higher surface gravities prevents dissociation \citep{Kirkpatrick1993}. However, hot Jupiters have lower surface gravities, and consequently photospheres at lower pressures (around 0.1 bar for WASP-18~b), thus their spectra should be depleted in water beyond 2700~K. Carbon monoxide is harder to thermally dissociate, and should be present for temperatures below 4000~K, as expected in WASP-18~b.
Furthermore, while the cross-section of water increases, the line contrast weakens at higher temperatures (e.g., \citealt{Burrows1997}). Hence, it is inherently harder to identify spectral features of water at high temperatures.
\par

Opacities from the negative hydrogen ion  H$^-$ are relevant at temperatures between 2500-8000~K (e.g., \citealt{Pannekoek1931, Chandrasekhar1945}, \citealt{Lenzuni1991}), hence they are important for very highly irradiated exoplanets (Figure~\ref{fig:spec}). Atomic hydrogen is produced through thermal dissociation of molecular hydrogen at these high temperatures (e.g., \citealt{Bell2017}), along with electrons from the metal ionization (see Figure \ref{fig:spaghetti}). Therefore, we argue that the hottest gas giants will show significant opacity from H$^-$ ions on their daysides. We study the importance of H$^-$ with planet mass and temperature in a companion paper (Parmentier et al. in prep.). \\

\begin{figure}
\begin{center}
\includegraphics[scale=0.45]{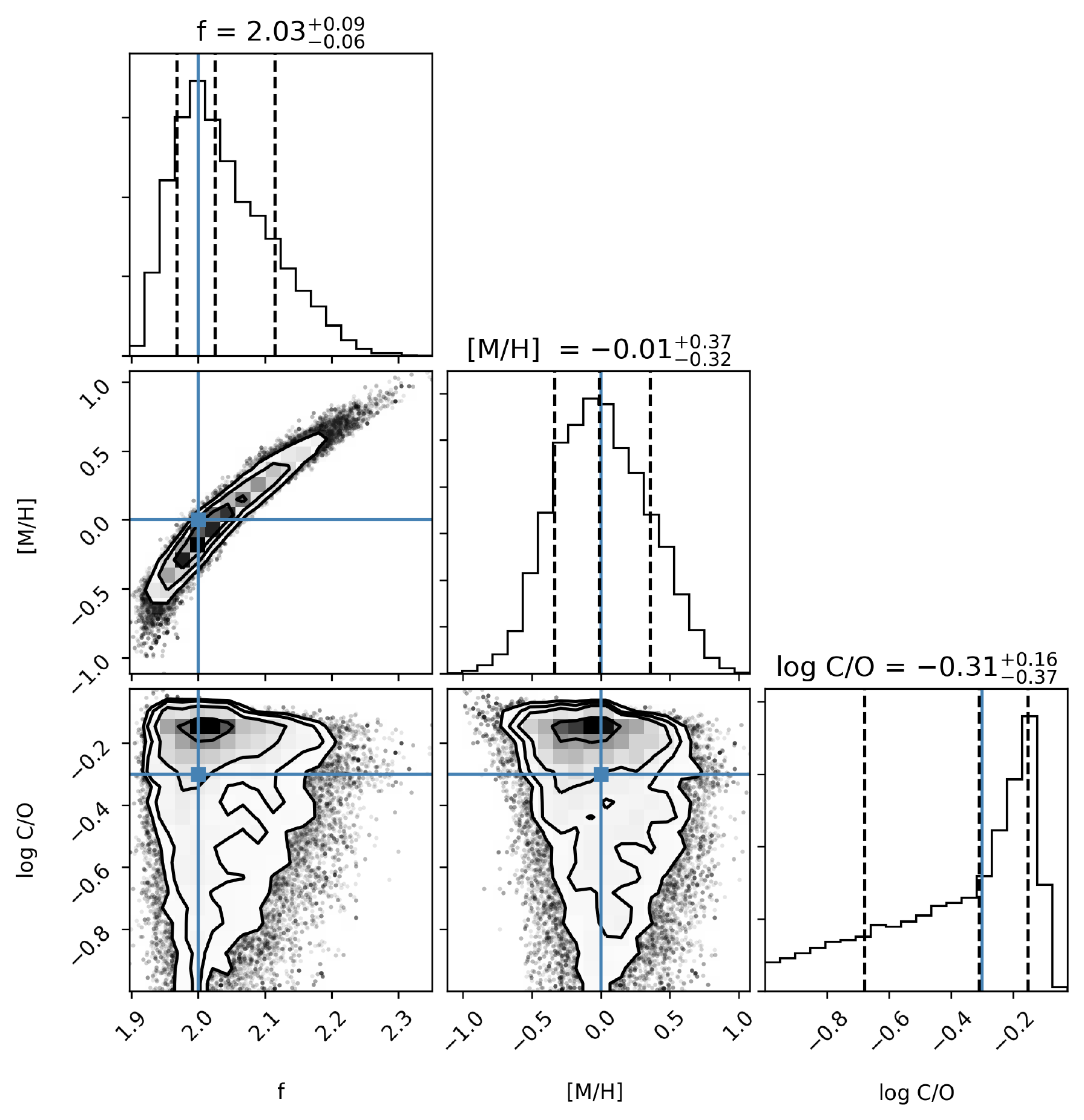}
\caption{Posterior distributions from the grid retrieval. Forward models calculated varying log(C/O), [M/H] and f (redistribution factor). The extracted metallicity and C/O ratio are consistent with solar values (plotted in blue), f is consistent with no day-night redistribution.} \label{fig:corner}
\end{center}
\end{figure}

\subsection{Atmospheric modelling including H$^-$ opacities and molecular dissociation}

We produce a newly developed cloud-free grid of 1D self-consistent radiative-convective-thermochemical equilibrium models to interpret the data (ScCHIMERA, Self-consistent CHIMERA, \citealt{Line2013}). We use the \cite{Toon1989} two-stream source function technique under the hemispheric mean approximation to solve for the infrared radiative fluxes at each atmospheric layer combined with a convective adjustment scheme in the deeper atmosphere. The incident stellar flux is modelled as pure attenuation at a disk averaged airmass of $1/\sqrt 3$. The molecular, ion, and condensate abundances are derived using the NASA CEA2 Gibbs-free energy minimization routine \citep{Gordon1994} given the elemental abundances scaled from \cite{Lodders2009} via the metallicity, [M/H], and carbon-to-oxygen ratio, C/O, while accounting for the depletion of elements due to condensate rain-out. We implement the line-by-line cross section database described in \cite{Freedman2008, Freedman2014} with in the correlated-K "resort-rebin" framework described in \cite{Lacis1991,Molliere2015} and \cite{Amundsen2016} at a constant resolving power of 100 between 0.3 and 200 $\mu$m. The code has been validated against analytic solutions and agrees to within 3\% and against the brown dwarf models of \cite{Marley2010}.  Bound-free and free-free opacities are taken from \cite{John1988} and \cite{Bell1987}, respectively.  The grid is parametrised with a scaling factor to the stellar flux ($0.75\leq f\leq 2.5$) to account for the unknown albedo and day-to-night heat transport (such that when f=1 there is complete day-night redistribution and when f=2 only the dayside re-radiates), the metallicity ($-1<$[M/H]$<2$), and carbon-to-oxygen ratio ($0.1<$C/O$<10$ with finer sampling between 0.75 and 2). Parameter estimation is performed over the grid using the EMCEE package \citep{emcee} via interpolation of the spectra along the grid dimensions, binned to the appropriate WFC3 and Spitzer resolution elements/profiles. The grid resolution is fine enough that interpolation errors are negligible. \par
We achieve a best fit with a reduced chi-squared of 2.0. We found that, when both H$^-$ opacities and dissociation effects were not included, our retrievals were pushed to high C/O in order to explain the lack of water features, as seen in other studies (e.g, \citealt{Sheppard2017}).
A pairs plot of the posterior distributions is shown in Figure \ref{fig:corner}. The metallicity is constrained to be solar ([M/H]~$=-0.01\pm 0.35$). A high C/O ratio is ruled out; we retrieve C/O$<0.85$ at $3\sigma$ confidence, also consistent with solar. Our retrieved value of f=$2.03\pm 0.08$ is consistent with minimal day-night redistribution expected for such a hot planet \citep{PerezBecker2013} and measured by \cite{Maxted2013}.

\subsection{WASP-18~b's atmospheric metallicity, composition, and thermal structure}

\begin{figure}
\begin{center}
\includegraphics[scale=0.5]{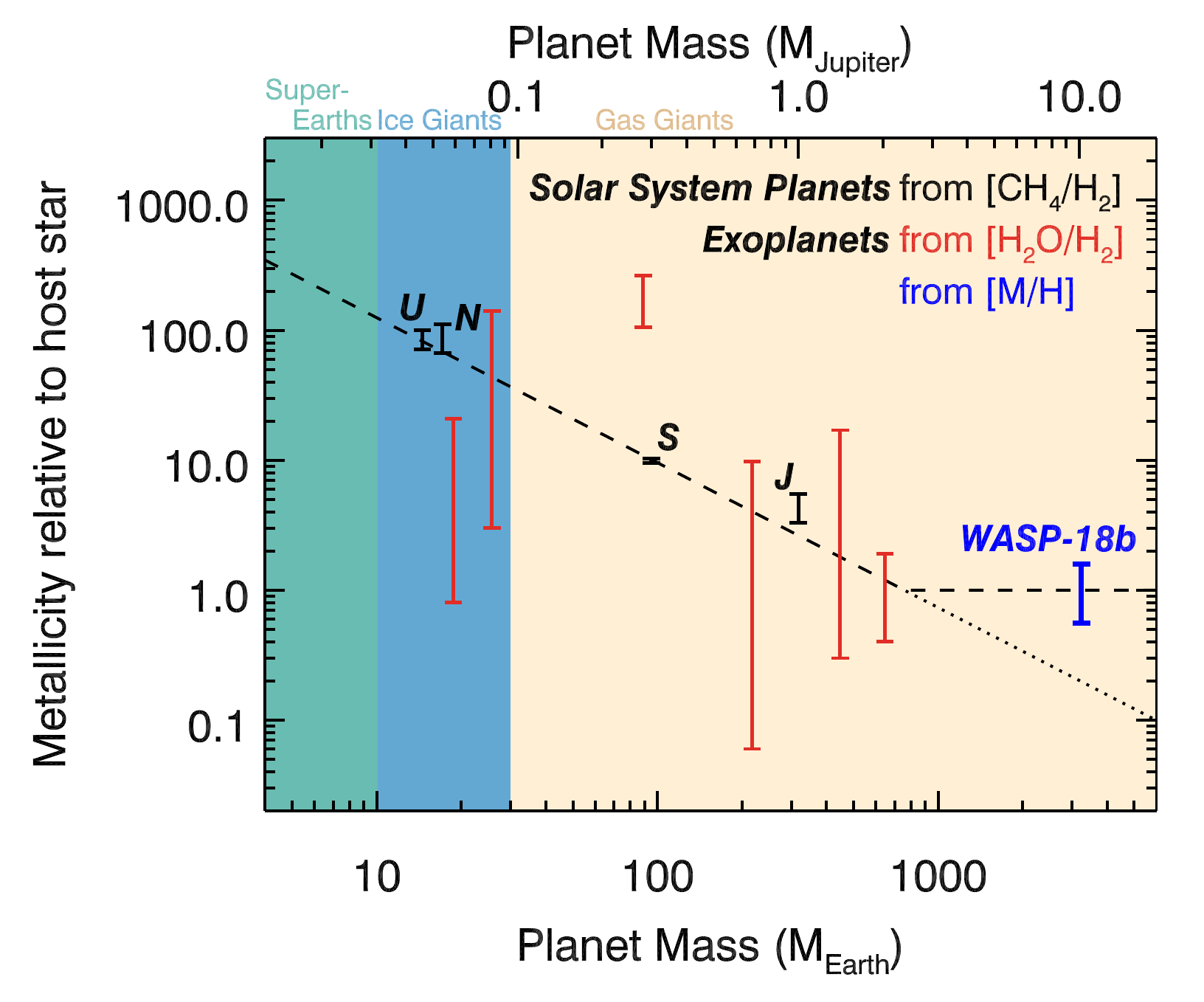}
\caption{Mass-metallicity plot of the solar system planets including known exoplanet metallicity estimates from measured water abundances (Mansfield et al. in prep. and references therein). For the most massive planets, the metallicity should not follow a log-linear relationship but should approach the metallicity of the host star, as seen in WASP-18~b where the stellar metallicity is 0.11$\pm$0.08 \citep{Torres2012}}.
\label{fig:mh}
\end{center}
\end{figure}

We compare the retrieved metallicity of WASP-18~b to the measured metallicities of solar system giants and exoplanets in Figure \ref{fig:mh} and show that WASP-18~b agrees with the expectation that the metallicities of the most massive planets should approach the metallicities of their host stars.

The tight constraint on metallicity, despite the absence of spectrally resolved molecular features, comes in part from the dependence of H$^-$ on metal fraction. The ionization of metals is the dominant source of free electrons that generate H$^-$ opacity in the atmosphere, and so there is a direct link between the H$^-$ continuum level and the abundance of metals. In particular, this is driven by the abundance of metals that are the dominant sources of free electrons (Na, K, and Ca; \citealt{Longstaff2017}). 
However, the complex relationship between the chemistry and the temperature structure as well as their joint effects on the spectrum make it difficult to attribute the retrieved metallicity solely to the H$^-$ continuum. 

The nominal self-consistent temperature pressure profiles (Figure \ref{fig:spaghetti}) show a thermal inversion with temperature increasing with altitude at pressures between 0.1-0.01 bar. The inverted T-P profiles are also required to fit the emission feature at 4.5 $\mu$m, due to CO and H$_2$O, as observed with Spitzer \citep{Nymeyer2011, Maxted2013}. 
This inversion in our models is caused by high altitude absorption of optical stellar light by TiO and VO, and reduced cooling due to the dissociation of water. Vertical cold trapping of TiO can act to remove this species from the atmosphere of hot Jupiters (e.g., \citealt{Desert2008}), but not for planets with equilibrium temperatures above $\sim$1900~K \citep{Parmentier2016}. Horizontal cold trapping could still remove inversions from gas giants with high surface gravities \citep{Parmentier2013,Beatty2017}, however we do not see this in WASP-18~b as our modelling favours an atmosphere with a TiO driven inversion.
In order to test the presence of the inversion we perform a second grid retrieval, but with the opacities of TiO and VO removed. Practically, this removes the temperature inversion for the cases where the C/O $< 0.8$. For higher C/O, oxygen-poor atmospheres are naturally depleted in TiO/VO so they can no longer be the source of the inversion. By comparing the Bayesian Information Criterion (BIC) we found that the models including TiO and VO were favoured ($\Delta$BIC=6.5) over those without, at odds with the retrieval by \citet{Sheppard2017}. Even though TiO/VO are present in our model, their features are not seen in the WFC3 bandpass as they are damped by the H$^-$ continuum while also being partially dissociated (seen in Figure \ref{fig:spec}). 
We finally compare the BIC between the best-fit model spectrum and the blackbody fit and find that the best-fit model to the combined HST/WFC3 and Spitzer/IRAC data is favoured over the isothermal atmosphere ($\Delta$BIC=12). Hence, our best fit favours a dayside model atmosphere with a solar metallicity and C/O ratio, and with a thermal inversion.

The abundance of water in the atmosphere is expected to be partially depleted by dissociation (see Figure \ref{fig:spaghetti}). While water is not dissociated at the pressure levels probed by the WFC3, dissociation of water higher in the atmosphere (below 0.1 bar) removes its emission feature at 1.4~$\mu$m. If dissociation were not present, the line centre of emission would originate from higher in the atmosphere where the temperature is greater.
We therefore attribute the the lack of water emission features both to an increased continuum opacity from H- and to decrease in line opacity by dissociation of water higher in the atmosphere. The final spectrum between 1.1-1.7~$\mu$m therefore appears featureless as it is a sum of broad, partially-depleted water emission at 1.4~$\mu$m and H$^-$ bound-free opacity between 1.1-1.4~$\mu$m (see also Parmentier et al. in prep.). 
However, the dominant effect in the case of WASP-18~b is the increased continuum opacity from H- over the thermal dissociation of water (brown contours, Figure \ref{fig:spaghetti}).

Another effect of water dissociation at low pressures is that it reduces the ability of the atmosphere to cool in this region \citep{Molliere2015}. Hence in our models, even though the partial dissociation of TiO reduces the heating of the upper atmosphere, the atmospheric cooling efficiency remains lower than the heating efficiency, producing a thermal inversion.

\section{Consequences for the family of very hot giant exoplanets}

Our results for WASP-18~b have consequences for the new family of very hot gas giants, where extrapolation from cooler planets can be misleading (WASP-33~b: \citealt{Haynes2015}, WASP-103~b: \citealt{Cartier2017}, WASP-18~b: \citealt{Sheppard2017}, WASP-121~b: \citealt{Evans2017}).
We find that the important impact of opacity both from H$^-$ formed from metal ionization and from the reduced abundance of species, including water, due to thermal dissociation is key to the interpretation of very hot gas giant atmospheres.
An evidence for this is that when including H$^-$ opacity, the metallicity and C/O of WASP-18~b are no longer super-solar, but drop to solar values. This is expected for typical formation scenarios of such a massive planet. Our result implies that the metallicity and C/O of other recently found metal-enriched very hot gas giants could also drop to solar values when H$^-$ opacity is considered.\par

Interestingly, almost all of the very hot gas giants probed so far are best explained with the presence of a thermal inversion. Indeed, the primary diagnostic of these thermal inversions is a flux excess at 4.5~$\mu$m \citep{Knutson2010}. This implies that the hottest exoplanets exhibit a common behaviour in their temperature structures, whose origin could be due to optical absorbers such as TiO/VO. Our modelling suggest that the WFC3 observations probe the region near the tropopause that is quasi-isothermal, and appears to produce blackbody-like spectra due to the combined effects of dissociation and H$^-$ opacity. Thus, we postulate that transition regions in classes of hot Jupiters could occur around temperatures near 2500~K (HAT-P-7~b, Mansfield et al. in prep.), below which H$^-$ opacity becomes less significant, and near 1800~K, below which TiO and VO condense. \\


We thank Christiane Helling and Mickael Bonnefoy for useful discussions, and Eliza Kempton for providing feedback on the manuscript. J.M.D. acknowledges that the research leading to these results has received funding from the European Research Council (ERC) under the European Union's Horizon 2020 research and innovation programme (grant agreement no. 679633; Exo-Atmos).
J.M.D acknowledges support by the Amsterdam Academic Alliance (AAA) Program. Support for program GO-13467 was provided to the US-based researchers by NASA through a grant from the Space Telescope Science Institute, which is operated by the Association of Universities for Research in Astronomy, Inc., under NASA contract NAS 5-26555. J.L.B. acknowledges support from the David and Lucile Packard Foundation. M.R.L. acknowledges the ASU A2D2 Saguaro and Agave computer clusters used for the bulk of the grid model computations.




\begin{table*}[h]
\begin{center}
\begin{tabular}{ | c | c | c | c || c | c | c | c |}
\hline
Wavelengths & Fp/Fs & Error & Model & Wavelengths & Fp/Fs & Error & Model\\
$\mu$m & ppm & ppm & ppm & $\mu$m & ppm & ppm & ppm \\
\hline
1.140-1.173 & 775 & 20 & 805 & 1.436-1.469 & 1131 & 21 & 1140 \\
1.173-1.206 & 874 & 20 & 870 & 1.469-1.501 & 1190 & 21 & 1187 \\
1.206-1.239 & 908 & 20 & 883 & 1.501-1.534 & 1237 & 21 & 1192 \\
1.239-1.271 & 908 & 19 & 917 & 1.534-1.567 & 1171 & 23 & 1221 \\
1.271-1.304 & 940 & 19 & 959 & 1.567-1.600 & 1205 & 24 & 1245 \\
1.304-1.337 & 989 & 20 & 986 & 3.6 & 3020 & 150 & 3081 \\
1.337-1.370 & 1050 & 20 & 1043 & 4.5 & 3850 & 170 & 3601 \\
1.370-1.403 & 1105 & 20 & 1077 & 5.8 & 3700 & 300 & 4043 \\
1.403-1.436 & 1141 & 21 & 1108 & 8.0 & 4100 & 200 & 4512 \\
\hline
\end{tabular}
\caption{Extracted secondary eclipse spectrum, including photometric Spitzer/IRAC points from \cite{Maxted2013} and \cite{Nymeyer2011} and best fit model from our grid-retrieval, convolved to the resolution of the data. Eclipse depths and 1$\sigma$ errors for HST/WFC3 were obtained using MCMC analysis on each of the spectroscopic light curves.}
\end{center}
\label{tab:spec}
\end{table*}


\end{document}